\documentclass[%
 preprint,
showpacs,preprintnumbers,
 amsmath,amssymb,
 aps,
 pra,
 longbibliography,
 lengthcheck,%
]{revtex4-1}

\usepackage[mathcal]{euscript}
\usepackage[greek,english]{babel}
\usepackage{cases}
\usepackage{enumerate}
\usepackage{xspace}
\usepackage{type1cm}
\usepackage{longtable}
\usepackage{graphicx}
\usepackage{epstopdf}
\DeclareGraphicsExtensions{.eps,.png}
\usepackage{dcolumn}
\usepackage{bm}
\usepackage{hyperref}
\usepackage{amsthm}
\usepackage{ifthen, paralist}
\usepackage{accents}
\usepackage[cal=boondoxo]{mathalfa} 
\usepackage{arcs}
\usepackage{color}

\newcommand{\bra}[1]{\mathop{\langle{#1}|}\nolimits}
\newcommand{\ket}[1]{\mathop{|{#1}\rangle}\nolimits}

\newcommand{\LR}[0]{{\mathop{\leftrightarrow}}}

\newcommand{\<}[1]{}
\newcommand{\pder}[2]{\mathop{\frac{\partial #1}{\partial #2}}}

\newcommand{\idx}[1]{\mbox{\scriptsize #1}}

\newcommand{\Tr}{\mathop{\rm{Tr}}}
\newcommand{\es}[1]{{\mathop{\mathtt{#1}}}} 
\newcommand{\ese}[1]{E_{\mathtt{#1}}} 
\newcommand{\esp}[1]{\ket{\mathtt{#1}}\bra{\mathtt{#1}}} 

\let\oldgather = \gather
\let\endoldgather = \endgather
\renewenvironment{gather}[0]{\par\nobreak\noindent\oldgather}{\endoldgather}
\let\oldalign = \align
\let\endoldalign = \endalign
\renewenvironment{align}[0]{\par\nobreak\noindent\oldalign}{\endoldalign}

\begin{document}

\newcommand{\APP}[1]{Appendix}

\title{Quantum optimal environment engineering for efficient photoinduced charge separation}

\author{Dmitry V. Zhdanov}
\email{dm.zhdanov@gmail.com}
\affiliation{Department of Chemistry, Northwestern University, 2145 Sheridan Road, Evanston, Illinois 60208-33113 USA}
\author{Tamar Seideman}
\email{t-seideman@northwestern.edu}
\affiliation{Department of Chemistry, Northwestern University, 2145 Sheridan Road, Evanston, Illinois 60208-33113 USA}

\begin{abstract}
The possibility to induce predetermined coherent quantum dynamics by controlling only the dissipative environmental parameters (such as temperature and pressure) is studied using the combined optimal control and environment engineering frameworks. As an example, we consider the problem of transforming an optically excited donor state into free charge carriers via intermediate higher-lying bridge state(s), with a view to solar energy conversion. In this context,  vibrational bath engineering allows to promote fast, directional charge transfer and to suppress recombinative losses.
\end{abstract}
\maketitle

\section{Introduction}
Understanding the charge and energy transfer in biological and synthetic molecular systems is essential for the development of efficient ecological energy sources and molecular electronics nano-devices \cite{2013-Zimbovskaya,2012-Kamat}. The dynamics of these systems belongs to a gray area between quantum and classical physics, where the efficacy and sustainability of the quantum coherent mechanisms is challenged by strong couplings with complex environments. Determining the optimal composition of coherent and incoherent couplings is challenging theoretically and experimentally. In particular, the role of the quantum coherences in fast and lossless charge transfer (CT) in biological photosynthetic complexes \cite{2007-Engel}  remains the subject of debate \cite{2014-Halpin}.

Tailoring the charge transfer system and its environment to be the most efficient in converting the incident incoherent solar radiation into free charge carriers is of great fundamental and practical interest. However, this problem lies outside the scope of the conventional quantum optimal control (OC) theory, which is primarily focused on guiding the dynamics of predefined systems by  shaping the external electromagnetic impacts (see e.g.~\cite{2012-Altafini,2012-Hoff} and references therein). At the same time,  new approaches to engineer the environment in dissipative dynamical systems for quantum state preparation and dynamical control were recently suggested \cite{2012-Muller,2012-Eremeev,2010-Pielawa,2012-Koga,2012-Marcos,2012-Tomadin,2009-Verstraete,2012-Murch,2006-Pechen,2006-Pechen-2,2013-Kastoryano} and experimentally demonstrated \cite{2011-Krauter}. In particular, they can be efficiently used to prepare highly entangled quantum states with high accuracy \cite{2012-Eremeev,2010-Pielawa,2012-Koga,2012-Marcos,2012-Tomadin,2009-Verstraete,2012-Murch} as well as to rapidly switch on and off  multiple decay channels with ultrafast precision \cite{2013-Kastoryano}. These studies together manifest that the external synergetic synthesis of coherent controls out of disorganized external energy inflow (as represented by laser generation) can be effectively substituted with the local self-organization process in the immediate quantum system's environment.
Moreover, the environmental controls potentially allow to steer the system into regions of the Hilbert space which are out of reach for optical coherent control \cite{2006-Romano,2007-Wu}.
These new achievements suggest considering the problem of effective charge transfer system design as a problem of integral quantum optimization of both junction properties (``Hamiltonian structure controls'' \cite{2012-Beltrani}) and reservoir parameters (``incoherent controls'' \cite{2007-Wu}).

In this paper we show that even relatively simple environment engineering allows to get remarkable benefit from constructive cooperation of resonant transitions and decay processes. Specifically, considering the case of charge transfer events,  we propose a robust and flexible mechanism to compensate the unfavorable intrasystem electronic arrangement when the charge transfer is originally strongly suppressed due to unfavorable energetics. Our results are based on a simple but very general
 3-electronic-state model of a junction in which the initial excited state of the donor decays into free carriers via an intermediate charge-transfer state. Despite its simplicity, the 3-state model is useful for describing single-molecule junctions \cite{2014-Kastlunger}, molecular bridges between quantum dots or molecules and semiconductors \cite{2013-Wang}, singlet-triplet exciton fission \cite{2013-Chan,2013-Jang} and conversion \cite{2012-Zhang}, as well as for semiqualitative analysis of the photosynthesis \cite{1996-Makri}. Its extensions, including multiple intermediate bridge states, also can account for the effect of quantum electronic interferences \cite{2011-Kocherzhenko,2014-Maggio}.

The paper is organized as follows. First, 
we present the mathematical details of our model and the optimization procedure. Then, 
we qualitatively and quantitatively analyze the outcome of numerical simulations. The concluding part 
contains the brief summary of the key results.

\section{The model description\label{@SEC02.-model_description}}
\begin{figure}[t]
\includegraphics[width=0.45\textwidth]
{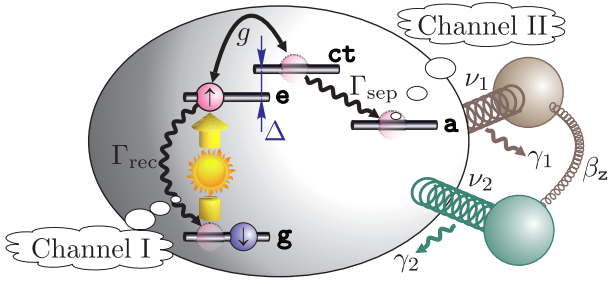}
\caption{The minimal electron transfer model.\label{@FIG.01}}
\end{figure}

In our model the donor excitation $\es{g}{\to}\es{e}$ by a random daylight photon evolves
with probabilities $P_{\mathrm{I}}$ and $P_{\mathrm{II}}$ towards
two channels (Fig.~\ref{@FIG.01}): (I) electron-hole recombination back to the ground state $\es{g}$ of donor, or (II) the two-step charge separation $\es{e}{\to}\es{ct}{\rightsquigarrow}\es{a}$. Here, oxidation of the donor is followed by slow incoherent decay of a charge transfer state $\es{ct}$ into a reduced state $\es{a}$ of acceptor. Our goal is to maximize $P_{\mathrm{II}}$ while suppressing the recombination.

All four involved electronic states are assumed to be weakly coupled with the modes of a cold Markovian phonon bath except for two modes $x_1$ and $x_2$ with frequencies $\nu_1$ and $\nu_2$. These two modes together with the mentioned electronic states form the basis $\ket{\es z,n_{x_1},n_{x_2}}$ of our reduced Hilbert space $\cal H$ where $\es z{=}\es{g,e,ct,a}$ and $n_{x_k}$ are the associated vibration quantum numbers.

In sequel we consider for simplicity the case of natural broadband illumination with constant spectral intensity in the relevant frequency range and neglect any sequential photon absorption effects which are unlikely 
on the realistic decay timescales in channels (I) and (II). At these conditions, the photon absorption at $t{=}0$ will lead to initial population of excited levels $\ket{\es e,n_{x_1},n_{x_2}}$ proportional to square of their overlap integrals with the 
equilibrium state $\rho_{\rm td}$.
The resulting excited state
\begin{gather}
\rho|_{t{=}0}{\propto}\sum_{n_{x_1},n_{x_2},n_{x_1}',n_{x_2}'}\ket{\es e,n_{x_1}',n_{x_2}'}\bra{\es e,n_{x_1},n_{x_2}}\times\notag\\
\Tr_{x_1,x_2}[\Tr_{\es z}[\ket{\es e,n_{x_1},n_{x_2}}\bra{\es e,n_{x_1}',n_{x_2}'}]\Tr_{\es z}[{\rho_{\rm td}}]]\label{model.-rho(0)[broadband]}
\end{gather}
will further evolve according to the 
Liouville equation:
\begin{equation}\label{model.-Liouville_equation}
\pder{\rho}{t}{=}{-}\frac{i}{\hbar}[\hat H,\rho]{+}{\cal L}_{\mathrm{diss}}[\rho].
\end{equation}

The Hamilton term of eq.~\eqref{model.-Liouville_equation} accounts for the charge transfer $\es{e}{\LR}\es{ct}$ with coupling constant $g$ and electron-phonon interaction with modes $x_1$ and $x_2$:
\begin{gather}
{\hat H}{=}g\ket{\es e}\bra{\es{ct}}{+}\sum_{\es z=\es{g,e,ct}}\esp z%
\left(\vphantom\int\right.\ese{z}{+}\sum_{k=1}^2\hbar\nu_k
\left\{\vphantom{^1}\hspace{-0.1cm}\right.(\hat a^{\dagger}_{k}{+}\alpha_{k,\es{z}})\times\notag\\
(\hat a_{k}{+}\alpha_{k,\es{z}}){+}\frac12\left.\vphantom{^1}\hspace{-0.1cm}\right\}
{+}
\beta_{\es{z}}
\prod_{k=1}^{2}\left\{\sqrt{\frac{m_k\nu_k}{2}}(\hat x_k{-}x_{k,\mathrm{z}})\right\}
\hspace{-0.2cm}\left.\vphantom\int\right)
{+}\mathrm{h.c.}\label{model.-Hamiltonian}
\end{gather}
Here $\ese{z}$ are electronic states energies, $\hat a_{k}$ is the phonon annihilation operator for the $k$-th mode in electronic state $\es e$, and the constants $\alpha_{k,\es{z}}{=}\sqrt{\frac{m_k\nu_k}{2\hbar}}(x_{k,\es{z}}{-}x_{k,e})$ are the scaled dimensionless shifts $(x_{k,\es{z}}{-}x_{k,\es{e}})$ of the vibration potential energy minima $x_{i,\es{z}}$ relative to ones in the state $\es e$. The last term on the right-hand side of eq.~\eqref{model.-Hamiltonian} describes vibrational intermode coupling with coupling constant $\beta_{\es{z}}$.

The dissipation part, ${\cal L}_{\mathrm{diss}}[\rho]$, in \eqref{model.-Liouville_equation} has form:
\begin{gather}
%
{\cal L}_{\mathrm{diss}}{=}\Gamma_{\rm sep}{\cal L}(C_{\es{a},\es{ct}})[\rho]{+}\Gamma_{\rm rec}{\cal L}(C_{\es{g},\es{e}})[\rho]{+}\notag\\
\sum_{k{=}1}^{2}\gamma_{k}{\cal L}(C_{\mathrm{vib},k})[\rho]{+}\sum_{\es{z}=\es{g,e,ct}}\Gamma_{z}{\cal L}(C_{\es{z},\es{z}})[\rho],\label{model.-dissipation_Liouvillian}
\end{gather}
where ${\cal L}(C)$ are conventional Lindblad terms:
\begin{gather}
{\cal L}(C)[\rho]{=}C\rho C^{\dagger}{-}\frac12(C^{\dagger}C\rho{+}\rho C^{\dagger}C),
\end{gather}
and
\begin{gather}
C_{\es z_2,\es z_1}{=}\ket{\es z_2}\bra{\es z_1};~~
C_{\mathrm{vib},k}{=}\sum_{\es z{=}\es{g,e,ct}}\esp{z}(\hat a_{k}{+}\alpha_{k,\es{z}}).
\end{gather}
The first two terms in \eqref{model.-dissipation_Liouvillian} represent the charge recombination $\es e{\rightsquigarrow}\es g$ and charge separation $\es{ct}{\rightsquigarrow}\es{a}$ decay channels with the corresponding decay rates $\Gamma_{\rm rec}$ and $\Gamma_{\rm sep}$. The remaining terms describe the thermal equilibration of the vibrational modes caused by weak Markovian collisional interaction with the cold bath $(kT{\ll}\hbar\nu_k)$. They include both inelastic contributions with the rates $\gamma_k$ and the elastic state-specific exponential transversal decays of the vibronic coherences with the rates $\Gamma_{\es{z}}$.

The dissipation dynamics in \eqref{model.-dissipation_Liouvillian} are described in terms of zero-order states $\ket{\es z,n_{x_1},n_{x_2}}$ and do not account for the effects of intermode and electronic couplings. The reliability of this approximation will be supported below by the numerical results. 

We will consider the ``worst-case'' situation shown in Fig.~\ref{@FIG.01} when 1) the charge transfer state $\es{ct}$ is above the state $\es e$: $\Delta{=}\ese{ct}{-}\ese{g}{>}0$; 2) electronic coupling is small: $g{\ll}\Delta$; 3) the electron decay rates are equal and slow: $\Gamma_{{\rm sep}}{=}\Gamma_{{\rm rec}}{\ll}g$. At these conditions and in absence of vibration modes $x_1$ and $x_2$ the value of $P_{\mathrm{II}}$ is small (at least, is evidently less than 50\%). Our goal is to increase the outcome of channel (II) by adjusting the phonon ``dressing'' which can be posed as an OC problem:
\begin{gather}\label{model.-OC_problem_settling}
\frac{P_{\mathrm{II}}}{P_{\mathrm{I}}}{\to}\max_{g,\nu_k,\Gamma_{\es z},\alpha_{k,\es z}\beta_{\es z}},
\end{gather}
where the parameters $g,\nu_k,\gamma_k,\Gamma_{\es z},\alpha_{k,\es z}$ and $\beta_{\es z}$ ($\es z{=}\es g,\es e,\es{ct}$; $k{=}1,2$) of vibrational environment are treated as ``controls''. The problem \eqref{model.-OC_problem_settling} was solved by numerical propagation of eq.~\eqref{model.-Liouville_equation} in conjunction with the genetic algorithm. The details are presented in \APP{@APP.01}.

\section{Results \label{@SEC04.-results}}
\subsection{Qualitative description}
\begin{figure}[b!]
\includegraphics[width=0.45\textwidth]
{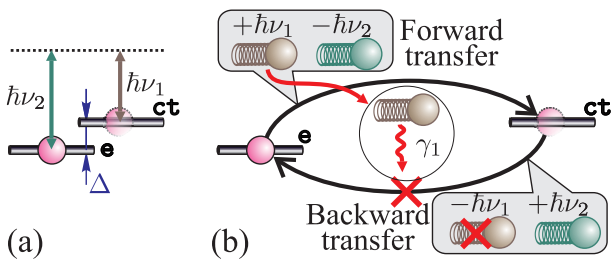}
\caption{(a) Scheme of optimal conditions for effective unidirectional electron transfer from the $\es e$ to $\es{ct}$ state (for $\kappa{=}1$). (b) Mechanism of population locking in $\es{ct}$ state: forward ${\es e}{\to}{\es{ct}}$ resonant transfer is possible whenever there exist vibrational quanta in $x_2$ mode; backward ${\es{ct}}{\to}{\es e}$ resonant transfer is locked just after relaxing the $x_1$ mode. \label{@FIG.02}}
\end{figure}
Let us first outline the rationale behind the optimal ``control'' parameters found in the simulations. The characteristic feature of all converged results (up to exchange of the vibration mode indices) is the following relations (Fig.~\ref{@FIG.02}a):
\begin{subequations}\label{results.-features}
\begin{gather}\label{results.-features_1}
|\Delta|{=}\hbar(\kappa \nu_2{-}\nu_1)~~~(\kappa=1,2,3....);\\
\gamma_1{>}0;~~~\gamma_2{=}0
\label{results.-features_2}
\end{gather}
\end{subequations}
The origin of relations \eqref{results.-features} for the case of $\kappa{=}1$ is clarified in Fig.~\ref{@FIG.02}(b). One can see that the quantum of energy from the mode $x_2$ ($x_1$) is required to support the forward (backward) resonant transition ${\es e}{\to}{\es{ct}}$ (${\es{ct}}{\to}{\es e}$). The relations \eqref{results.-features_2} imply that the vibrational energy resource for forward transitions is conserved whereas the excitations in $x_1$-mode ``leak away'' into the bath modes. Thus, if the geometry of the vibrational potential energy surface is such that the initial optical excitation is accompanied by strong vibrational excitation of the $x_2$-mode, then the forward transitions are resonant and fast while the backward transfer is off-resonant and slow. Thus, the population will be effectively locked in the $\es{ct}$ state until all the optically induced energy supplies in $x_2$-mode will be consumed. 

\subsection{Numerical illustration}
As an example, we present the numerical results for the case of $\frac{\Delta}{\hbar}{=}30$ and $\Gamma_{\rm sep}{=}\Gamma_{\rm rec}{=}0.01$. In order to get a clue into the excited state dynamics it is instructive to consider the model case in which the incident light uniformly populates the cluster of $N_1{\times}N_2$ vibronic states:
\begin{equation}\label{method.-rho_0}
\rho|_{t{=}0}{=}\frac{1}{N_1N_2}\sum_{n_{x_1}{=}0}^{N_1}\sum_{n_{x_2}{=}\delta N_2}^{\delta N_2{+}N_2}\ket{{\es e},n_{x_1},n_{x_2}}\bra{{\es e},n_{x_1},n_{x_2}}.
\end{equation}

All the identified solutions of the OC problem \eqref{model.-OC_problem_settling} with initial excited state \eqref{method.-rho_0} satisfy eqs.~\eqref{results.-features}, where the value of $\kappa$ depends on the choice of summation limits in \eqref{method.-rho_0}. Specifically, the optimal parameters for choice $\delta N_2{=}3$, $N_1{=}11$, and $N_2{=}13$ correspond to $k{=}3$: 
$\alpha_{1,{\es{ct}}}{=}0.5$, $\alpha_{2,{\es g}}{=}{-}2.69$, $\alpha_{2,{\es{ct}}}{=}{-}3.02$, $\nu_1{=}9.28$, $\nu_2{=}13.09$, $\frac{g}{\hbar}{=}2.35$, $\gamma_1{=}0.26$, $\Gamma_{\es{e}}{=}\Gamma_{\es{ct}}{=}\gamma_2{=}0$, $\beta_{\es g}{=}0$, $\beta_{\es e}{=}{-}0.11$, $\beta_{\es{ct}}{=}0.17$.

\begin{figure*}[t!]
\includegraphics [width=0.8\textwidth]
{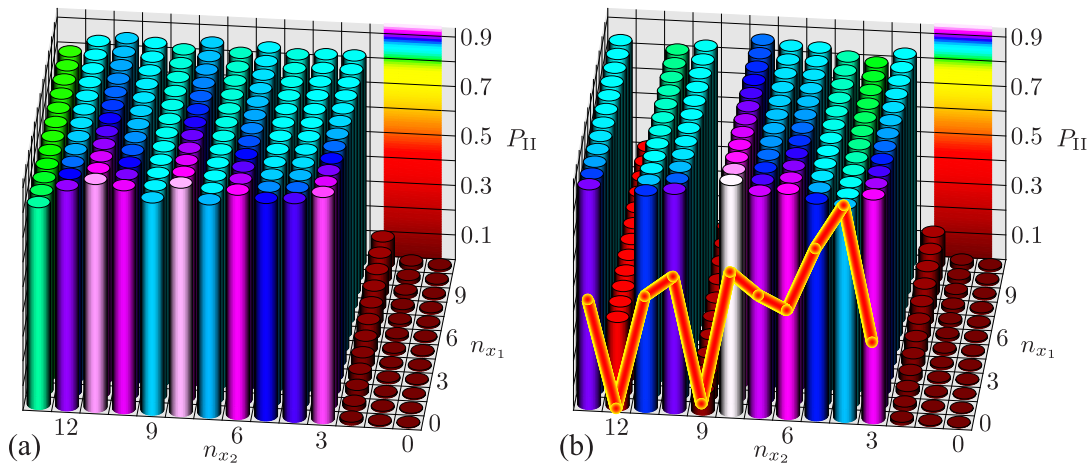}
\caption{(a) The probability of various excited states $\ket{\es e,n_{x_1},n_{x_2}}$ to evolve towards the channel (II) for optimized system parameters (see the text). (b) The same dependence with disabled vibrational couplings: $\beta_{\rm e}{=}\beta_{\rm ct}{=}0$. The nodes of red line show the relative absolute values of the overlap Franck-Condon integrals for $\ket{\es e,n_{n_{x_1}},n_{x_2}}{\to}\ket{\es{ct},n_{n_{x_1}},n_{x_2}{-}\kappa}$ transition.\label{@FIG.03}}
\end{figure*}

The contributions of the individual levels $\ket{\es e,n_{x_1},n_{x_2}}$ to the overall performance $P_{\rm II}\simeq 0.85$ of this solution are detailed in Fig.~\ref{@FIG.03}a. One can see that the probability of reaching the acceptor state increases abruptly once $n_{x_2}{\geq}\kappa$ (i.e{.} once the resonant ${\es e}{\to}{\es{ct}}$ transition becomes allowed) and slightly decreases with increasing $n_{x_1}$ (recall that $x_1$-quanta ``promote'' the backward transitions ${\es{ct}}{\to}{\es{e}}$). We have checked that the result is virtually insensitive to the presence of initial coherences between states $\ket{\es e,n_{x_1},n_{x_2}}$.
At the same time, the performance degrades substantially in the absence of the intermode couplings $\beta_{\es e}$ and $\beta_{\es{ct}}$ (Fig.~\ref{@FIG.03}b). Comparison with Fig.~\ref{@FIG.03}a shows that these couplings enhance the resonant transfer rate in the cases when the transitions are suppressed by small values of Franck-Condon overlap integrals.

\begin{figure}[t!]
\includegraphics 
{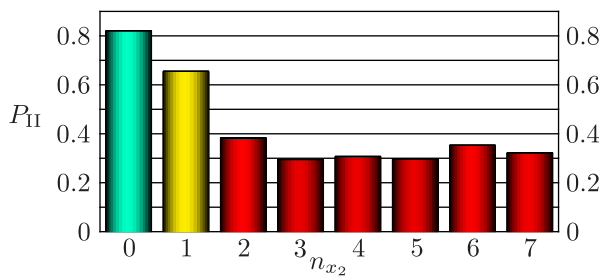}
\caption{Overall efficacy of the light-induced charge separation as a function of the vibrational state $\ket{\es g,n_{x_1},n_{x_2}}$ of the parent state. The excitation is considered within the broad spectrum approximation. Only the dependence on the $n_{x_2}$ vibrational quantum number is shown for fixed $n_{x_1}{=}0$; the dependence on $n_{x_1}$ is negligible. The ``controls'' were optimized for launching from the ground vibrational state of $\es g$ with $n_{x_1}{=}n_{x_2}{=}0$. \label{@FIG.04}}
\end{figure}

The efficacy of charge separation is also very high in the case of broadband incoherent solar energy absorption (see eq.~\eqref{model.-rho(0)[broadband]}). 
In this case the maximal performance $P_{\rm II}\simeq 0.85$ can be achieved with $\alpha_{1,{\rm g}}{=}0$ at low temperatures, when the equilibrium vibronic state $\rho_{\rm td}$ coincides with the ground state $\ket{\es g,0,0}$. However, the results of calculations summarized 
in Fig.~\ref{@FIG.04} allow to predict substantial efficacy losses at higher temperatures when the thermal excitation of mode $x_2$ is non-negligible.

\begin{figure}[t!]
\includegraphics 
{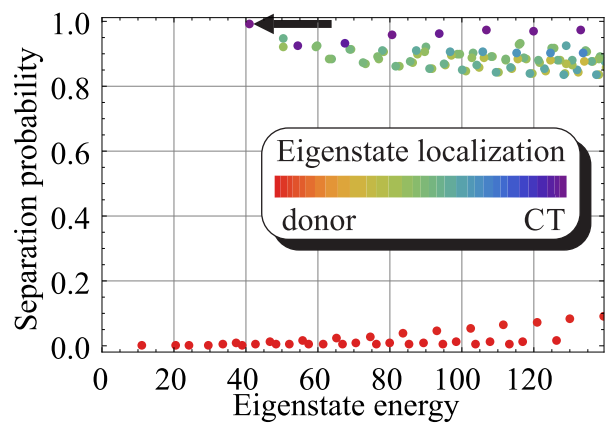}
\caption{Probability of reaching the outcome (II) starting from the different eigenstates of the  Hamiltonian \eqref{model.-Hamiltonian}. The localization of each eigenstate is indicated by color codes as clarified in the inset.\label{@FIG.05}}
\end{figure}

Fig.~\ref{@FIG.05} provides a complimentary way to elucidate the charge separation mechanism in terms of the evolution of the individual eigenstates of the Hamiltonian \eqref{model.-Hamiltonian}. One can see that the major ``feeders'' of channel (I) are strongly localized on the donor. More detailed analysis shows that all of them are composed of such zero-order states $\ket{\es e,n_{x_1},n_{x_2}}$ that $n_{x_2}{<}\kappa$, so that the forward transfer $\es e{\to}\es{ct}$ is energetically forbidden. In contrast, the vast majority of the remaining excited states predominantly evolve towards the acceptor side regardless of their structure and localization. Due to thermal relaxation, the populations of these states at times $\frac1{\gamma_2}{\ll} t{\ll}\frac1{\Gamma_{\idx{sep,rel}}}$ will be accumulated in the lowest energy ``drain'' state of the upper group (indicated by an arrow in Fig.~\ref{@FIG.05}). This state has pronounced charge transfer character and is essentially decoupled from the donor side. For this reason, its decay dynamics can be accurately represented within the zero-order basis approximation. These observations confirm the 
applicability of the model assumptions made above.

We remark, interestingly, that the ``drain''-type decoupled eigenstates that sustain the radiationless decay within the strongly mixed vibronic systems, have been discovered in realistic molecular models and studied in context of optimal control \cite{2004-Sukharev}. Their remarkable properties have been related to the phenomenon of scars of periodic orbits \cite{2005-Sukharev}.

\section{Summary and conclusion}
In this work we proposed an approach to (incoherently) controlling daylight-induced charge separation, hence a potential route to solar energy conversion. The control  mechanism is
\begin{itemize}
\item universal. It can be applied without any substantial change to stimulate a variety of other types of unidirectional quantum dynamics (e.g. energy transfer, proton transfer etc.) in different donor-bridge-acceptor systems;
\item effective, irreversible and fast. In fact, it is capable to enforce even counterintuitive sustainable population flow from the lower-energy to higher-energy electronic state;
\item essentially insensitive to the spectral characteristics of the incident light and hence potentially well-suited to harvesting broadband solar energy.
\end{itemize}

The substitution of lasers with readily available cheap energy sources is a fascinating challenge with obvious ramifications. Based on the results of this work we propose the following concepts for achieving this goal:

\begin{itemize}
\item The concept of a vibrational reservoir as a dual-use resource to achieve  unidirectional quantum dynamics: The associated coherent mechanisms promote the resonant forward $\es{e}{\to}\es{ct}$ transfer while the incoherent decay channels are responsible for population locking in the target state $\es{ct}$. From the energy standpoint, this concept exploits the fact that optical excitation, in general, brings both the electronic and the vibrational subsystems into non-equilibrium, and subsequent evolution of the latter imposes coherent feedback on the former. As a result, all the essential quantum coherences are created locally without use of external coherent energy resource.

\item The concept of marrying quantum optimal control methods with environment engineering to determine the characteristics of the reservoir which will lead to a desired outcome. In other words,  the quantum optimal control is put in the framework of the traditional chemistry methods, so that not the laser fields but the temperature, pressure, parameters of solvent, chemical composition etc. become the key quantum control parameters.
\end{itemize}

One of the interesting questions that this study opens is whether the proposed mechanism is also exploited by natural light-harvesting complexes. Specifically, it provides the natural framework for converting any initial coherences which can be encoded in the certain vibrational modes by laser excitation into electronic coherences similar to observed in the recent 2D-spectroscopic experiments \cite{2007-Engel,2014-Halpin}. The analysis and verification of this possibility would carry us beyond the scope of the present paper.

\begin{acknowledgments}
We thank Dr. Sameer Patwardhan for stimulating discussions and feedback.
\end{acknowledgments}

\appendix

\section*{Appendix: The optimal control of charge transfer\label{@APP.01}} 
\subsection{Numerical estimation of the probabilities \texorpdfstring{$P_{\idx{I}}$}{Lg} and \texorpdfstring{$P_{\idx{II}}$}{Lg}}
The definition of $P_{\idx{I}}$ and $P_{\idx{II}}$ implies that these quantities are feasible physical parameters only until the electron dynamics remains confined in the Hilbert space $\cal H$. In practice, this corresponds to timescales $T$ of order $T{\sim}\min{}(\Gamma_{\idx{sep}},\Gamma_{\idx{rec}})^{{-}1}$. The key simplifications (one-electron picture, absence of sequential absorption and recombination $\es a{\to}\es g$, Markovian phonon bath approximation) embedded into the Liouville equation~\eqref{model.-Liouville_equation} are also justified for this time window. However, the character of these simplifications ensures that the populations of states $\es g$ and $\es a$ formed on the timescales of $T$ will remain ``frozen'' during further ``unphysical'' propagation of eq.~\eqref{model.-Liouville_equation}. In particular, this means that:
\begin{gather}
\rho|_{t{=}\infty}{=}P_{\idx{I}}\varrho_{\idx{I}}{+}P_{\idx{II}}\varrho_{\idx{II}},\label{method.-rho(t=infty)}
\end{gather}
where $\varrho_{\idx{I}}{=}\ket{\es g,0,0}\bra{\es g,0,0}$ and $\varrho_{\idx{II}}{=}\ket{\es a,0,0}\bra{\es a,0,0}$ are the linearly independent steady eigenstates of the total Liouvillian ${\cal L}[\odot]{=}\frac{{-}i}{\hbar}[\hat H,\odot]{+}{\cal L_{\idx{diss}}}\odot$: ${\cal L}[\varrho_{\idx{I,II}}]{=}0$. Denote as $\Psi_{\idx{I}}$ and $\Psi_{\idx{II}}$ the associated normalized right eigenvectors: $\Psi_{\idx{I}}{\cal L}[\odot]{=}0$; $\Tr[\varrho_{{\xi}}\Psi_{{\xi}}]{=}1$; $\Tr[\varrho_{{\chi{\ne}\xi}}\Psi_{{\xi}}]{=}0$ ($\xi,\chi{=}$I,II). The last three relations lead to the following identity:
\begin{gather}\label{method.-linear_system_for_Psi_d}
\Tr[\Psi_{\idx{II}}\left({\cal L}\odot{+}\varrho_{\idx{II}}\Tr[\hat O\odot]+\varrho_{\idx{I}}\Tr[\odot]\right)]{=}\Tr[\hat O\odot],
\end{gather}
where $\hat O$ is an arbitrary Hermitian operator such that the superoperator in the round brackets is full-rank (e.g{.} $\hat O{=}\varrho_{\idx{II}}$). Multiplying \eqref{method.-rho(t=infty)} by $\Psi_{\idx{II}}$ and taking trace gives:
\begin{gather}
P_{\idx{II}}{=}\Tr[\Psi_{\idx{II}}\rho|_{t{\to}\infty}]{=}\Tr[\Psi_{\idx{II}}\rho|_{t{}0}];~~~P_{\idx{I}}{=}1{-}P_{\idx{II}},\label{method.-P_I,P_II-value}
\end{gather}
where the fact of time-independence of $\cal L$ is used.

Eqs.~\eqref{method.-linear_system_for_Psi_d} and \eqref{method.-P_I,P_II-value} constitute the complete system for determining $P_{\idx{I}}$ and $P_{\idx{II}}$. In simulations we represented eq.~\eqref{method.-linear_system_for_Psi_d} in appropriate operator basis as a system of linear equations $L_{i,j}\Psi_j{=}o_i$ relative to $\Psi_j$ and then solved it using the conjugate gradient method with a diagonal Jacobi preconditioner $M{=}{\rm diag}\{L_{1,1},L_{2,2},...\}$ \cite{BOOK-Saad}.

\subsection{Solution of the extremal problem \texorpdfstring{${P_{\mathrm{II}}}/{P_{\mathrm{I}}}{\to}\max$}{Lg}}

The approximate extrema of the objective functional $\frac{P_{\mathrm{II}}}{P_{\mathrm{I}}}{\to}\max$ were searched using the genetic algorithm. In our calculations we used the genotype which includes 11 parameters: $g,\nu_1,\nu_2,\gamma_1,\gamma_2$, $\Gamma_{\es e}$, $\Gamma_{\es{ct}}$, $\alpha_{k,\es e}$, $\alpha_{k,\es ct}$, $\beta_{\es e}$ and  $\beta_{\es{ct}}$ as well as its extended 13-parameter variant where the vibrational frequencies $\nu_1$ and $\nu_2$ are allowed to be the functions of electronic state $\es z$. We found, however, that the maximal efficacy is achieved in absence of dependences $\nu_{1,2}(\es z)$. The optimization was performed on the population of 25 individuals. Each subsequent generation includes the best 34$\%$ of parents, another 33$\%$ are obtained via random mutation of the 20$\%$ of genes in seeding individuals and the rest result from crossover of $20\%$ of genes between two seeding individuals. The seeding individuals were randomly selected among the best 50$\%$ of parents merged with extra 3-4 randomly generated candidate solutions.

After reaching substantial convergence the best solution was further refined using the standard gradient descent method.

\end{document}